# Designing a Linked Data Migrational Framework for Singapore Government Datasets

# Technical Report

**Aravind Sesagiri Raamkumar, Muthu Kumaar Thangavelu, Sudarsan Kaleeswaran & Christopher S.G. Khoo**
Wee Kim Wee School of Communication and Information
Nanyang Technological University
Singapore 637718

## ABSTRACT

The subject area of this report is Linked Data and its application to the Government domain. Linked Data is an alternative method of data representation that aims to interlink data from varied sources through relationships. Governments around the world have started publishing their data in this format to assist citizens in making better use of public services. This report provides an eight step migrational framework for converting Singapore Government data from legacy systems to Linked Data format. The framework formulation is based on a study of the Singapore data ecosystem with help from Infocomm Development Authority (iDA) of Singapore. Each step in the migrational framework has been constructed with objectives, recommendations, best practices and issues with entry and exit points. This work builds on the existing Linked Data literature, implementations in other countries and cookbooks provided by Linked Data researchers. iDA can use this report to gain an understanding of the effort and work involved in the implementation of Linked Data system on top of their legacy systems. The framework can be evaluated by building a Proof of Concept (POC) application.

# 1. INTRODUCTION

The world contains vast amount of digital information in different forms – facts, graphs, charts and maps. Organizing, consuming and publishing these huge volumes of data to create new information and understand the existing information better is becoming increasingly complex. There is a huge opportunity for linking data and reusing to form a rich network of knowledge, by standardizing in a common format. In the future, the links can enable intelligent chain of transactions on the web without human intervention and also for easy comparison with similar information in the web of data. 'Linked Data' is one such standard establishing a common format and allows publication in the open data cloud to link with other data sets that enable reuse. The study focuses on implementation of such a data standard for Infocomm Development Authority (iDA) data.gov.sg (DGS) portal. The portal publishes the governance data provided by different ministries and agencies in Singapore. The current data eco system is studied and a customized Linked Data migration framework is designed by selectively using the conversion tools and techniques in different stages of migration. The migration steps are elaborated by using population trends and urban authority's sites data sets. The potential application after migration is demonstrated with the inter-linking of these data sets across domains within the data.gov.sg platform.

The cookbooks, guidelines and recommendations provided by the Linked Data research communities are helpful in understanding the general steps and tools required in converting and publishing government data in Linked Data format. Governments that are new entrants in adopting Linked Data need a tailored data migration framework specific to their local data ecosystem. Existing case-studies recommend different tools for migration thereby creating an additional overhead of tool evaluation before implementation. Therefore, it is necessary to recommend the best tool based on evaluation in the local context.

The current study aims to build a Linked Data migrational framework that could be used by iDA and Singapore Government agencies to publish their data sets in the form of Linked Data. The framework build process is based on the metadata and specifications provided by iDA and government agencies. The explanation of each step in the framework is given with two data sets provided by Urban Redevelopment Authority and Department Of Statistics (described in Table 1). Each step in the framework would comprise of objectives, recommendations and issues identified by the project team based on careful evaluation. The current study focuses on linking the internal data sets. Additionally, it aims to provide recommendations on a few use-cases that leverage the utility of external Linked Data.

| Data set | Agency | Category | Data type | Existing Applications |
|---|---|---|---|---|
| Resident Population by DGP Zone/ Subzone and Age Group, Type of Dwelling, Ethnic Group | Department of Statistics | Population and Household Characteristics | Textual | Singstat Time Series (STS), charts on singstat themes |
| Sites Sold by URA - Details | Urban Redevelopment Authority (URA) | Housing and Urban Planning | Textual | OneMap view of site location |

**Table 1. Selected Datasets for Linked Data Conversion Study**

# 2. LITERATURE REVIEW

The aim of Linked Data is to create a web of actionable data, standardized with a common format on the internet. Linked Data is a standard for achieving semantic web. Semantic web can be called a systematic way of discovering knowledge by representing data with relationships (Berners-Lee, Hendler & Lassila, 2001).

Berners-Lee propagated simple principles for linking data on the web. It starts with identifying the resources and assigning Uniform Resource Identifier, URI (different from Uniform Resource Locator URL that identifies as well as locates). The identifiers should follow HTTP protocol so that it can be accessed on web browsers. The publishing standard of the data can be in a common format called RDF (Resource Description Framework). The common data format can be linked with external identifiers (URI) and they can discover related things (Shadbolt, Hall & Berners-Lee, 2006). This initiative of publishing data in a 'Linked Data format' became widely popular in libraries and government organizations that contain data from different domains and formats, with a need to be reorganized in a common format for easy search, retrieval of information and to enhance the productivity of data with mash ups, map based and real time applications. (Halb, Raimond & Hausenblas, 2007)

Linked Open Data2 (LOD2)[1] project is studied for understanding the existing Linked Open Data stack of products, frameworks and processes. The project aims to accelerate the implementation of Linked Data across the globe with open source tools and tutorials. The start of the Linked Data movement spurred the release of new data sets highlighted by the Linked Open Data cloud maintained by Comprehensive Knowledge Archive Network (CKAN) registry. US and UK governments have already realized the benefits by releasing selective data sets in the Linked Data format in the portals data.gov[2] and data.gov.uk[3] respectively. Well-defined relationships between these datasets and ready-made applications guide public's daily activities related to transport, business and other needs.

The best practices and existing methodologies for open governmental data conversion to Linked Data standards are studied with the cook books provided by Bizer, Heath, Idehen & Berners-Lee (2008), Villazón, Vilches, Corcho & Gómez-Pérez (2011) and Hyland & Wood (2011).

The steps in the migrational framework are chosen after a thorough analysis of the existing frameworks for Government Linked Data migration. The conversion framework suggested in (DiFranzo, Graves, Erickson, Ding, Michaeli, Lebo, Patton, 2011) has been implemented in US. The steps are Name, Retrieve, Adjust, Convert, Enhance, and Publish. Though the framework has a good coverage of core migration steps, it is not well grounded with the emphasis on specifications and object modeling and its main focus is on faster conversion. The migration steps suggested in (Hyland & Wood, 2011) are comprehensive and it is taken as a base for developing a customized framework for Singapore. The framework works systematically with Modeling, Naming URIs, Vocabulary reuse, Ontology mapping and heads towards core conversion to RDF standards, publication and licensing. RDF conversion step describes the trade-off for choosing triplification and modeling approach and also the choice of service providers. This model served as a base for the designed migration framework.

The third migration approach suggested in (Villazón-Terrazas, Vilches-Blázquez, Corcho, Gómez-Pérez, 2011) has a similar process to the previous framework. It starts with codifying the system specifications that involves URI design and modeling, identifying the vocabularies for reuse and creating new ontologies on the top, generation of RDF triples with data transformation (standardization), cleansing, linking and discovering relationships between data items of government dataset, publication (data and meta data) and exploitation by effective linking to the Linked Open Data cloud and creating intelligent mash ups and reasoners in the application layer.

---

[1] LOD2 Project http://lod2.eu/BlogPost/9-press-release-lod2-project-launch.html
[2] US Government Open Data data.gov
[3] UK Government Open Data data.gov.uk

## 3. METHODOLOGY

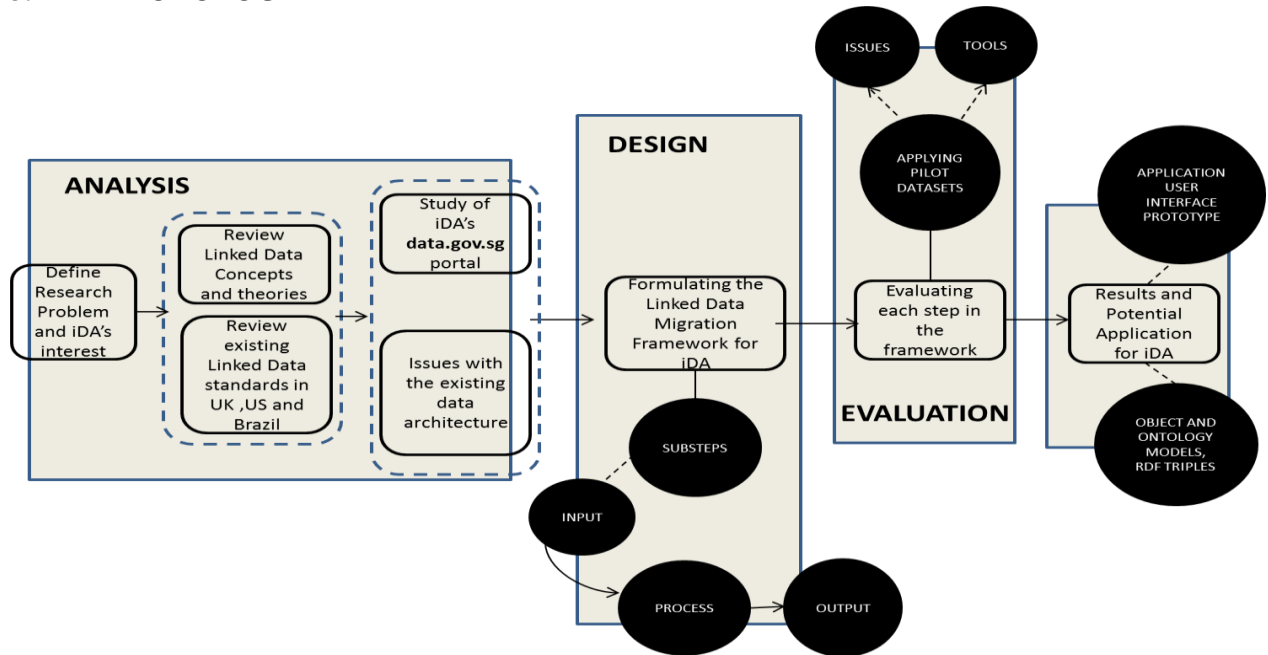

**Figure 1: Research Methodology – Linked Data Migrational Framework Design for iDA**

The activities in the project are classified under three phases – Analysis, Design and Evaluation.

### 3.1 Analysis

The phase deals with the collection of required information from the source parties iDA, SLA and NIIT through face to face meetings followed by investigation of the current public data ecosystem of Singapore Government.

Singapore Government publishes its public data in a portal data.gov.sg (DGS) as a part of the eGov2015 initiative. Infocomm Development Authority of Singapore (iDA) maintains the portal. iDA collects data from different government agencies (Chee Hean, 2011). The data portal aims to meet Singapore public's data needs and to establish a co-creative environment and efficiently connect to public. The key actors identified in the data.gov.sg system are agencies providing data, data consumers (public), Infocomm Development Authority to interface & publish the data and the application developers exploiting the data with useful applications for the public. The governance data is currently provided by seven ministries and 50 agencies or statutory boards.

As in most countries, Ministries in Singapore govern over agencies thereby forming a natural hierarchy. For example, Ministry of Transport governs over Public Transport Council, Land Transport Authority, Civil Aviation Authority, Maritime and Port Authority. It has been identified that hierarchies have not been considered in the data.gov.sg system's user interface and hence becoming a downside for the data consumers in associating data from the same domain under different end points. The existence and flow of data in the current data eco system is studied with iDA's system design documents preceded by discussions with Infocomm Development Authority (iDA), Singapore Land Authority(SLA) and iDA IT vendors NIIT technologies staff working on the data.gov.sg (DGS) and OneMap projects.

| Artifacts Analyzed |
|---|
| 1. Source Data forms |
| 2. Data transfer and storage in data.gov.sg (DGS) system |
| 3. Data upload and download specifications |
| 4. Destination data forms |
| 5. Existing APIs for data consumption |
| 6. Selected data sets for Linked Data conversion study |
| 7. Current data consumption points for the selected data |

**Table 2. Singapore Government Data Ecosystem – Artifacts Analysis**

### 3.1.1 Source Data Forms
The data provided by the agencies exists either in textual or spatial form. The textual data from different agencies are retrieved from SG DATA database maintained by Department of Statistics (DOS). The spatial data from different agencies are maintained in OneMap database maintained by Singapore Land Authority (SLA).

### 3.1.2 Existing APIs for Data Consumption
DGS also provides Application Programming Interfaces (API) to facilitate easy data consumption in customized applications for the public and developer communities. OneMap API provided by Singapore Land Authority, Traffic-related APIs from Land Transport Authority, Tourism-related APIs from the Singapore Tourism Board, Environment-related APIs from the National Environment Agency and Library-related data feeds and web services from National Library Board are the APIs provided by DGS.

### 3.1.3 Issues identified with the Existing DGS System
The current issues identified with existing DGS system are described in this section. These issues are as per the findings of the research team.

1. Agencies do not provide raw data to Infocomm Development Authority. Aggregated report data is split into X dimensions representing columns, Y dimensions representing rows and data points representing cells. These fields are provided in an XML file and sent to Infocomm Development Authority on a periodic basis. There is no separate master data file. The hierarchy in master data dimensions is not explicitly set or provided. Therefore, a mechanism to identify the master data and the relationship between different levels in the master data dimensions needs to be devised.

2. In data.gov.sg (DGS) database, the references are made with Metadata id where the data is dumped in just two columns. There is an inherent lack in meaning about the retrieved data and its attributes. In most cases, there are issues with granularity (For example, location is expressed in streets and roads in Urban Redevelopment Authority 'sites for sale' dataset, whereas it is in planning areas in the Department of Statistics 'population trends' data set).

3. There is a conflict of Metadata standards in Textual and Spatial data, DGS and OneMap database (for example, 'category' has different set of values and classification in data.gov.sg and OneMap).

4. There is no master data management system in place right now that standardizes the dimension values across agencies. Standardization is required to link common data in the data sets. This might be a complex task due to the different versions of master data values in a single data set and also across data sets.

5. The current Governmental Data ecosystem of Singapore provides multiple end points to the users such as API, web services and files (Figure 2) making interlinking a complex process in application development.

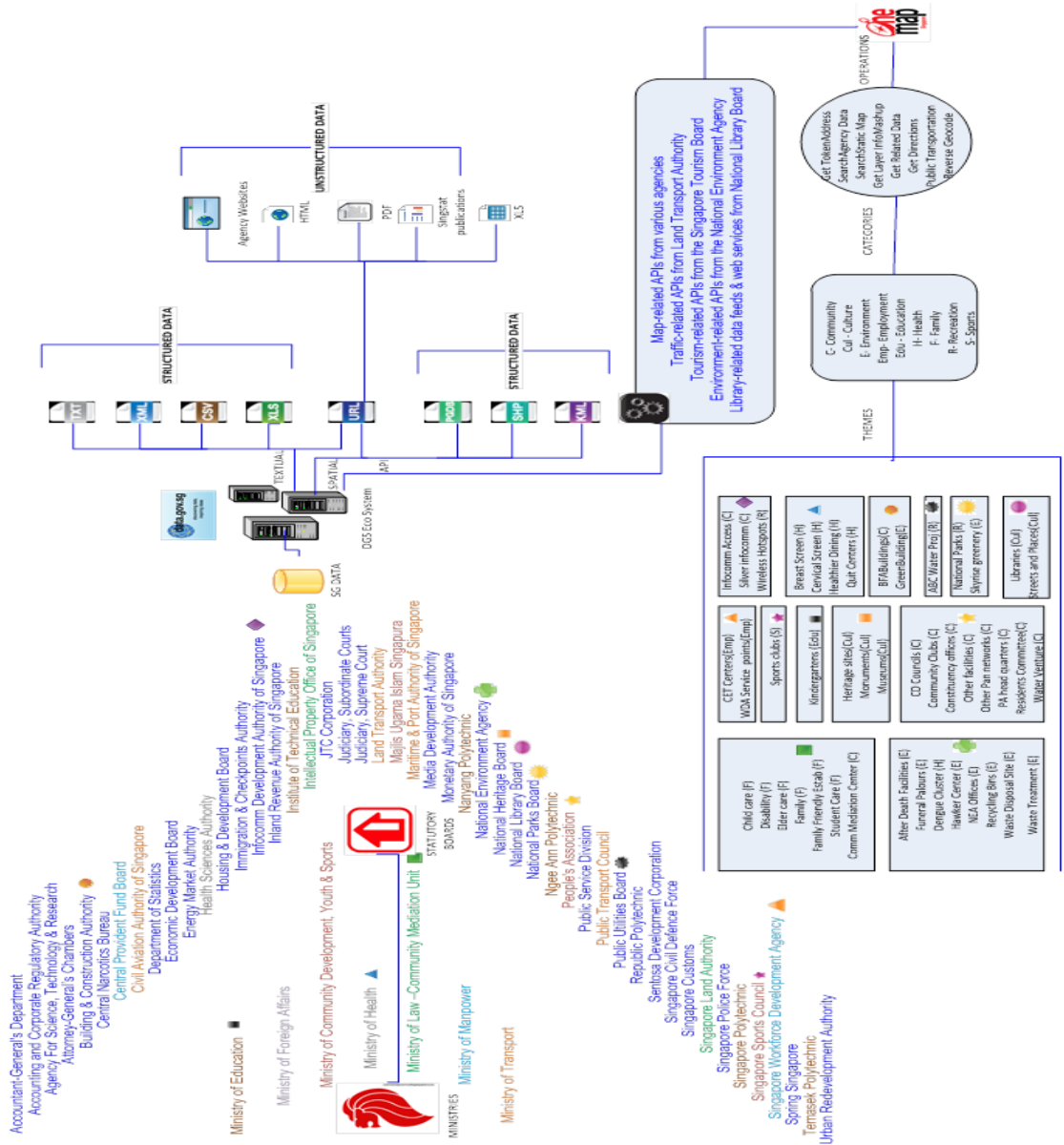

**Figure 2. Singapore Government Data Ecosystem – Comphrehensive Analysis Results**

Understanding the data.gov.sg and OneMap systems without the technical specifications of OneMap system and sample data from data.gov.sg tables from iDA was a challenge.

The study of existing migration frameworks, cookbooks and Linked Data implementations in data.gov.uk and data.gov (UK and US open government portals respectively) is done. The pilot datasets have been selected to demonstrate the capability of linked data to interlink domains (Urban planning and population demographics) on an application for the business executive intending to purchase a new site listed by Urban Redevelopment Authority (URA). The datasets from Urban Redevelopment Authority (URA) and Department of Statistics (DOS) have been used to demonstrate the application. There is a possibility of missing some sources as the analysis output is not signed off by IDA.

### 3.2 Design
The construction of the framework is done in the Design phase. The devised steps are to encompass every activity involved in the migration with clearly identified entry and exit points in each step. Identification of best tools from the inventory has been done for evaluation in the next phase. This selection is based on popularity, utility and compatibility measures. Similar exercise has been carried out for listing the standard vocabularies that could be used by IDA during the Ontology Modeling step.

The key objective is to design a Business Process style representation for each step with sub steps with respective input, process and output. An inventory list of the Linked Data tools that can be used in each step of the framework has been prepared by collecting information from literature and other web sources.

Recommendations on specific vocabularies applicable to each government agency will be one of the best fit deliverables in the migration design phase. However, only pilot datasets of two agencies (URA & DOS) were associated with vocabularies. Details about the design phase are furnished in the Migrational framework section of the report.

### 3.3 Evaluation
The devised framework along with the recommended tools is applied on the pilot data sets in the Evaluation phase. Three prominent RDF Conversion tools (Google Refine, RDF Views and RDF Sponger) have been used for migration of the data to Linked Data format. The issues that iDA could encounter during the actual implementation is identified by this process of evaluation of the tools and steps.

The key objective is to provide a launch pad for IDA to get a first-hand view of the existing Linked Data tools in the market. The mitigation measures for the issues identified in each step have also been discussed. Evaluation criteria, usage and issues encountered are summarized with appropriate screenshots. This is intended to save time for iDA during the actual implementation.

On the downside, only few tools were evaluated by the project team. Final steps after RDF creation provide less scope for evaluation in current research. There may be more issues encountered with the increase in volume of data. Details about the evaluation phase are furnished in the Migrational framework section of the report.

## 4. MIGRATIONAL FRAMEWORK
The migrational framework has been devised based on the results of the SG Data eco system study and Linked Data literature review. The steps are sequential with clearly distinct entry and exit points. IDA can use the framework in an incremental/prototype mode so as to get an early view of the final output. The following subsections represent each step that has been explained using a Business Process perspective i.e. Input, Process and Output followed by Issues, Recommended Tools and Suggestions (as in Figure 3).

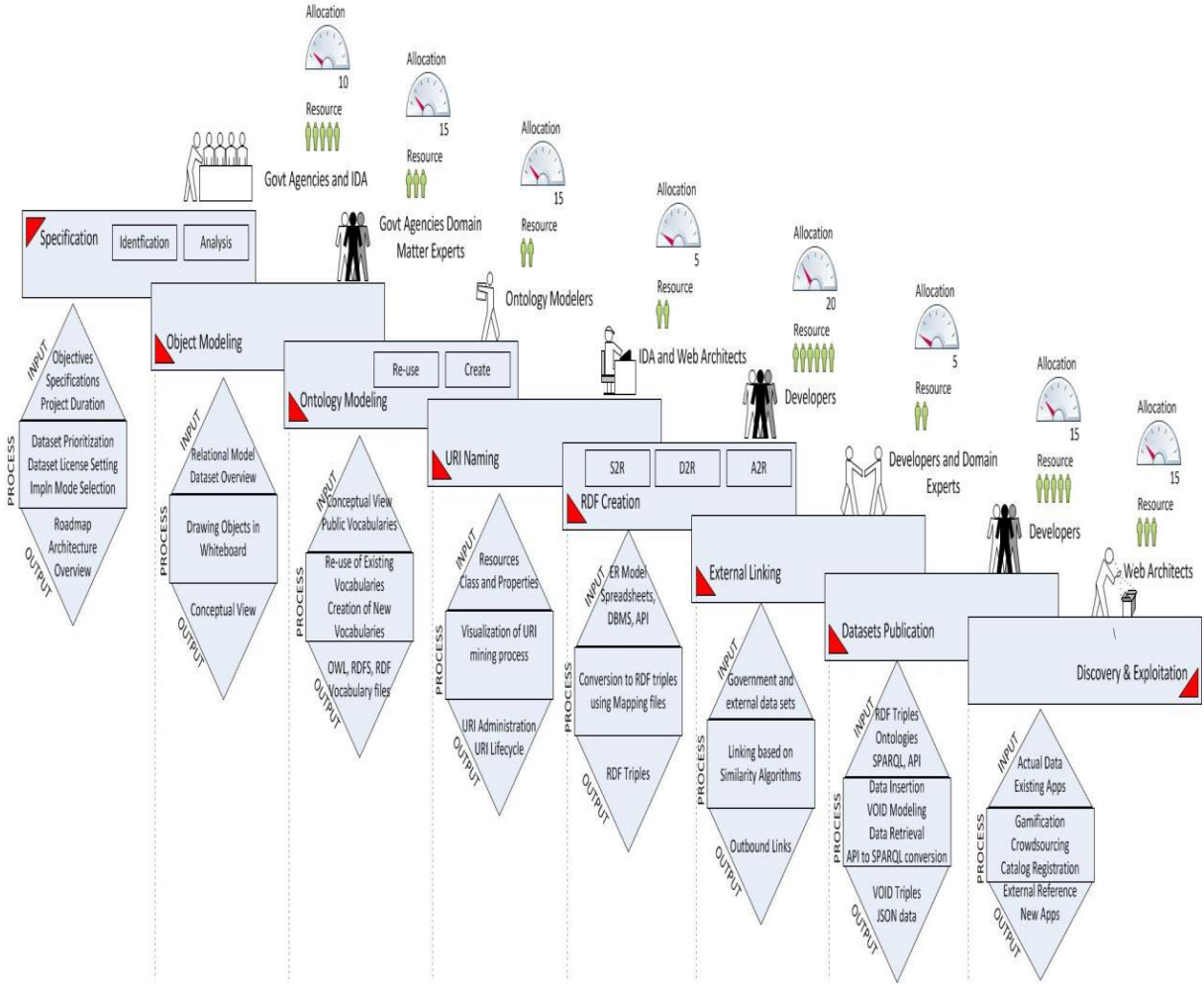

**Figure 3. Migrational Framework**

## 4.1 Step 1 - Specification
Specification is a de-facto step for any project at enterprise or government level. The nature of this step varies at the Program (IDA's Linked Data Program) and Project (Agency Linked Data implementation).

The activities in the steps are decided based on the study of the system and database specification documents of the data.gov.sg (DGS) to understand the logical model connecting the tables that is used to generate data sets and interpret the required specifications considering the steps in the existing frameworks available in (Hyland & Wood, 2011) and (Heath & Goodwin, J. 2011). There are two sub-steps in the process: - Identification (Program level) and Analysis (Project level)

### 4.1.1 Sub-step 1: Identification Activities at Program Level

*Pre requisites for Linked Data implementation*
The key security concerns that are to be addressed include providing API access keys and tokens for users. The key cost concerns include Linked Data hosting and use of commercial enterprise Linked Data conversion software such as Virtuoso. The mode of Linked Data implementation is to be chosen based on the needs of the publisher (Infocomm Development Authority in our case)

*Implementation Planning*
The high level architecture of the implementation is to be designed. The data sets need to be classified as per the order of implementation. This is decided by evaluating the usefulness of data sets and the scope for interlinking with other data sets across domains by Subject Matter Experts in agencies. A specific Linked Data conversion potential can be assigned based on the possible in-links and out-links to the dataset. Based on this potential, a data set can be classified into high, medium and low (More the potential is, more priority needs to be given during the conversion process). The Linked Data conversion of low potential data sets can be optional.

The implementation coverage can have the possible options - Agency specific implementation (Specific agencies like National Environmental Agency or Department of Statistics choose to convert their data with a higher Linked Data potential) and Application specific implementation (The data used by specific applications in the catalogue are converted). This can be decided based on the T&A (Time and Allocation) budget possible options.

Another important consideration is the license to be set for the resulting data sets. This often depends on the source of data. The different types of license are based on the sensitivity and confidentiality of data. The access rights and means of data access are decided based on license (Refer Table 3).

| **Types of License** |
| --- |
| The Open Database License (ODbL) |
| Open Data Commons Attribution License |
| The Creative Commons Licenses |

Table 3. License Types for iDA

For example, the first two open type licenses can release data on the web directly as RDF Dumps whereas the third might require restricting RDF access directly to users and provide SPARQL endpoint (A means to access data in Linked Data standard) or Application Programming Interface (API).

### 4.1.2 Sub-step 2: Analysis Activities at Project Level
The system analysis and design documents (including database) are collected. Information about identified data sets in Data.Gov.Sg (DGS) including Metadata, schema design and Entity Relationship (ER) models are studied with iDA system integration and database specifications.

| Sub-step | Input | Process | Output |
| --- | --- | --- | --- |
| **Identification (Program)** | • Budget<br>• Program Duration<br>• Objectives | • Data set prioritization<br>• Data License Assignment<br>• Implementation mode selection | • Roadmap<br>• Architecture<br>• Implementation Coverage |
| **Analysis (Project)** | • Specifications<br>• Relational models<br>• ER models | Resource and schema mapping | Overview of data set |

Table 4. Business Process Overview of Specification Step

### 4.1.3 Identified Issues
The improper licensing of data sets can lead to illegal usage and leakage of sensitive and critical information (e.g. crime and abortion rates in Singapore). Present and future public data providing government Information System projects need to be planned with the possibility of having a downstream option with Linked Data. Improper planning leads to changes in architecture and affect project timelines.

### 4.1.4 Recommended Tools
There are no Linked Data specific tools recommended for this step due to its strategic nature. The project team used tools such as MS Project, MS Visio for their planning and designing purpose.

### 4.1.5 Other Suggestions

IDA has to select the Linked Data implementation before proceeding to the subsequent steps. The project team recommends IDA to use the "Linked Data URI only" method initially so that the URI management becomes stable followed by the "Linked Data+RDF" method (Refer Table 5).

> **Method1 – Linked Data only (just URI linking),** the second quickest mode of implementation enables re-use of instances across agencies. It can be used for facts publishing. Disadvantage is the additional work needed to be done for getting actual data about the URIs from the user and system. IDA can use this method to test the URI management process of Centralized (DGS) vs. Decentralized (Agency controlled).
>
> **Method 2 - Linked Data + RDF,** the most comprehensive use of Linked Data requires IDA to do additional work to create separate RDF files for each resource. Therefore, it needs extensive storage. IDA can use this method to completely realize the benefits of Linked Data and Semantic Web standards. A decision to use this mode can be taken after evaluating the first prototype/POC application
>
> **Method 3 – Linked Data for files (URIs for files only),** the quickest mode of implementation can help in Search and Retrieval operation by offering rich metadata. There is a lack of interlinking between data sets and the applicability of this method for files and not at data level. IDA can use this method to improve the discovery of file based data sets in DGS by tagging files.

**Table 5. Possible Linked Data Implementation modes for iDA**

### 4.2 STEP 2 - OBJECT MODELING

Object Modeling is aimed at clearly depicting the objects & their relationships in the data sets and is to be done with help from domain experts of the respective government agencies. Knowledge attained from the overview of the data sets should suffice to the requirements of this step. Objects from the data sets (e.g. Location, Building Type, Race and Area) should be drawn on a whiteboard and lines should be drawn to indicate the relationships between the objects (e.g. Locations fall under Zone, population falls under demographic data etc.). The entire exercise is to be done with coverage on mind and not based on the futuristic uses of the applications. It is recommended to normalize the model in 3NF (Third Normal Form) before starting the exercise.

The project team drew the conceptual view of the data set objects in the whiteboard (as in Figure 4). The key learning from this experience was the relative ease in identifying the use of common classes between the data sets (e.g. the use of Location object) and brainstorming the relationship between objects from different data sets.

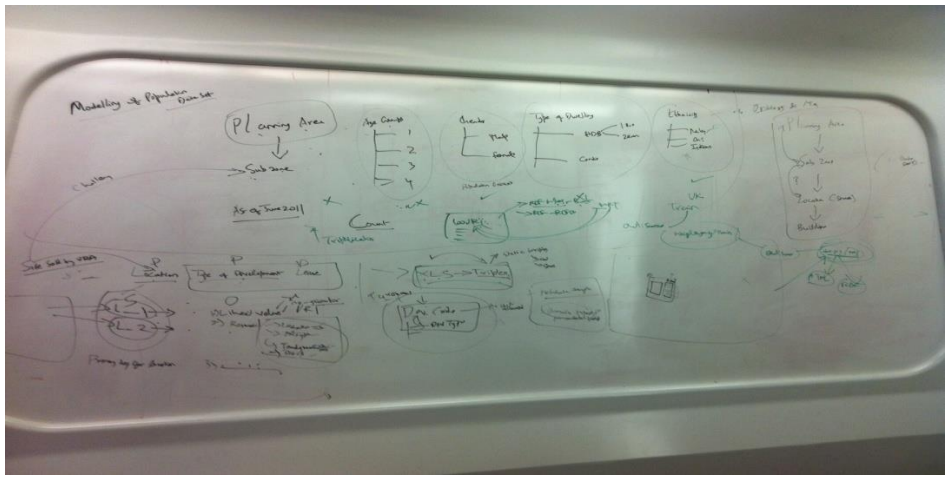

**Figure 4. Object Modeling done by the Project Team**

| Sub-step | Input | Process | Output |
|---|---|---|---|
| **Object Modeling**(for individual data sets and between data sets) | <ul><li>Overview of data sets</li><li>Relational model of data sets</li></ul> | Drawing objects and linking them in whiteboard | Conceptual view of all objects in the system with their interconnections |

**Table 6. Business Process Overview of Object Modeling Step**

### 4.2.1 Identified Issues
There might be a tendency to apply high abstraction over some objects and concurrently apply high granularity to others objects in purview. For example, middle levels in the zone hierarchy can be missed and road level data can be mistaken as classes at the same time. Hastened closure of this exercise can result in poor design. The output diagram should be signed off by domain experts of the concerned government agencies

### 4.2.2 Recommended Tools
Concept map is apt tool for object modeling. The project team recommends the usage of whiteboard for this purpose even though electronic drawing of the objects and their relationships can also be done.

### 4.3 STEP 3 - ONTOLOGY MODELING
The interlinking of data is achieved by modeling the abstract entities close to the real world into classes, sub classes, object and data properties and instances. This is called ontology or sometimes called a concept vocabulary. The efficiency of these ontologies lies in reusing rather than recreating and constraining the use of data with a significant degree of structure. (Siricharoen, 2007). In the current data.gov.sg (DGS) system, the data from these data sets are used in separate contexts as singstat visualizations in charts and Urban Redevelopment Authority (URA) site map. A common use case scenario across domains is furnished in Table 7.

> Consider an industrial entrepreneur intending to buy a site from Urban Redevelopment Authority (URA) to start his new industry, might be interested to look at other aspects of the site along with the data provided by URA like plot ratio and development type, to validate purchase option thoughts instantaneously, say the population trends - age group, ethnicity, living standard, employability and even the location trends – transportation available to the site location, nearest expressway, distance to the nearest port (air and sea) for shipment.

**Table 7. Use case problem statement to realize the benefits of Ontology across domains**

The object models (a concept map) of the data sets developed in the previous step was used to study the higher level relationships within data. In the population trends data, the following hierarchies and relationships are identified. The population demographics have one of the examining factors (Age, race, location, and gender or employment status). The location is a region, planning area or a street location. The dwelling types are HDB, Condominiums, Private Flats, or Landed Property. One of the object properties of planning area is 'has Site Location'. This property carries an inverse relationship with the property of site location - 'under Planning Area'. The URA site for sale data is organized based on the sites (site location as classes). The sub class identified is the type of development code, in turn has a sub class, 'Type of development allowed'.

The common thing that looked inter-connectable in these datasets is the information based on 'location'. Ontology is developed to establish the relationship between the data sets based on the location data. The existing vocabulary Geonames is reused. The data in different granularity levels is a challenge for standardization. Urban Redevelopment Authority (URA) sites for sale data contain location at specific site location level only. Department of Statistics (DOS) Population trends data set contains both at planning area and specific site location levels. Geo names vocabulary contain at postal code level.

A standardized location vocabulary is derived with the site hierarchy provided by Urban Redevelopment Authority (URA) and mapped to both the data sets extended to postal code levels and reusing the geonames vocabulary (URI). The country has regions Central, West, East, North and North East. It has 12 urban planning areas (for e.g. Clementi, Boon lay, Jurong West). The urban planning area has site location like Pioneer Road North, Woodlands Avenue. The site location has a postal code (e.g. 628462 for pioneer road north).

Networking ontologies and extending the relationships across domains improves the machine reasoning and intuitive information retrieval from the user's perspective as seen as above.

| Sub-step | Input | Process | Output |
|---|---|---|---|
| **Ontology Modeling** (for individual data sets and between data sets) | • Individual data sets(URA and DOS data sets)<br>• Object models developed in the previous step | Creating explicit relationships with classes, sub classes, object and data properties on the domain specific information. | Ontology linking urban planning and population domain data sets |

**Table 8. Business Process Overview of Ontology Modeling Step**

### 4.3.1 Issues
Resolving conflicting vocabulary and Metadata standards in data.gov.sg (DGS) and OneMap as explained in the issues with the existing system could be a challenge. The existence of different levels of granularity in data sets. For example, Location is expressed in different forms in different instances – urban planning area, specific site location or postal codes. Building a common location vocabulary to network these different granular data can be used to uniformly drill down or roll up the location information chain in applications with a single version of truth.

### 4.3.2 Recommended Tools
Neon Tool Kit is recommended for design, storage, version maintenance and deployment of large ontologies. This also has an option to import the standard reusable vocabularies automatically and is best suited at enterprise level for its ease of use in interfacing vocabularies and visualizing the ontology graphs with defined links and hierarchies in a single shot as flash video or image. (Dzbor et al., 2005). Protégé, an open source tool can be used for prototyping.

## 4.4 STEP 4 - URI NAMING
Uniform Resource Indicators (URIs) are the building blocks of RDFs. URIs provide names to the resources in the system. IDA has to decide the URI administration method at the start of this step (as in Table 9).

> **Selection of URI Administration Modes**
> 1.) Maintained centrally in the DGS platform (resultant URIs will start with http://data.gov.sg/)
> 2.) Maintained by individual agencies (resultant URIs will start with http://ura.gov.sg or http://sla.gov.sg).
> 3.) Maintained externally by third party platforms such as Kasabi (resultant URIs will start with http://data.kasabi.org).
> **Key Decision Areas**
> 1.) Identifying the base URI (based on the output of URI administration decision step)
> 2.) Design the taxonomy of the URI i.e. structure of the URIs.

**Table 9. URI administration modes and key decision areas**

Most government Linked Data systems use the TBOX and ABOX approach for differentiating classes and their instances. TBOX refers to the resource URIs (http://data.gov.sg/resource/) and ABOX refers to Ontology URIs (http://data.gov.sg/ontology/)

The project team is of the view that URI minting process would be challenge for IDA if it goes for a piece-meal approach i.e. scattered implementation of Linked Data. URI construction should be done as a holistic process with good foresight on the future prospects of the system. IDA can arrange a workshop with all government agencies to educate and look for ideas pertaining to the representations that the agencies would want. The output of the workshop would be the taxonomy of the URI structure. The project team provides some recommended URI patterns for implementation as in Table 10.

| ABOX | TBOX |
|---|---|
| http://data.gov.sg/ontology/Ministry/ | http://data.gov.sg/ministry/MOH |
| http://data.gov.sg/ontology/Agency/ | http://data.gov.sg/agency/SLA |
| http://data.gov.sg/ontology/SiteLocation/ | http://data.gov.sg/location/pioneer_road_north |
| http://data.gov.sg/ontology/Race | http://data.gov.sg/race/chinese |
| **Dataset URIs** | |
| Dataset ID | URAstaticfile001 |
| Dataset | http://data.gov.sg/dataset/ URAstaticfile001/ |
| Class | http://data.gov.sg/terms/class/URAstaticfile001/sitesforsale |
| Property | http://data.gov.sg/terms/property/URAstaticfile001/time |
| Row 1 | http://data.gov.sg/dataset/URAstaticfile001/1 |
| Row 1 - A generic column | http://data.gov.sg/dataset/URAstaticfile001/1/columnName |

**Table 10. URI Patterns for Singapore Data**

| Sub-step | Input | Process | Output |
|---|---|---|---|
| URI Naming | • List of resources<br>• List of Classes and Properties | Visualization of URI minting process from Linked Data tools and usage perspective | • URI Administration method<br>• Base URI<br>• Taxonomy of URI structure<br>• URI lifecycle |

**Table 11. Business Process Overview of URI Naming Step**

### 4.4.1 Identified Issues
Usage of different vendor tools can hamper URI naming process on the situation of the tools not supporting custom coding of URIs. This can be averted by careful selection of tools. Dead links can affect applications that use the URIs. This can be monitored by using Vapour, a URI validation tool recommended by Pedantic Web Group.

### 4.4.2 Recommended Tools
The URI naming strategies of (Hendler, 2012), (Vila-Suero, 2012) and (Davidson, 2009) provide good guidelines. Pubby is a Linked Data tool that can be used for external URI dereferencing.

### 4.5 STEP 5 - RDF CREATION
RDFs are the building blocks of Linked Data systems. The successful execution of the earlier steps in the framework leads to this fruition step that involves the conversion of data from source systems to Linked Data format. The project team split the conversion process into three categories Static files conversion (S2R), Relational Database Management System (RDBMS) data conversion (D2R) and API data conversion (A2R). Examples of the different source formats are given in the below Table12 and Figure 5.

| Type | Example of Singapore data sets |
|---|---|
| S2R | URA Site for Sales, Singstat's Population Household Characteristics |
| D2R | DGS tables |
| A2R | OneMap API, myTransport API, NLB web services |

**Table 12. Conversion Types with Examples**

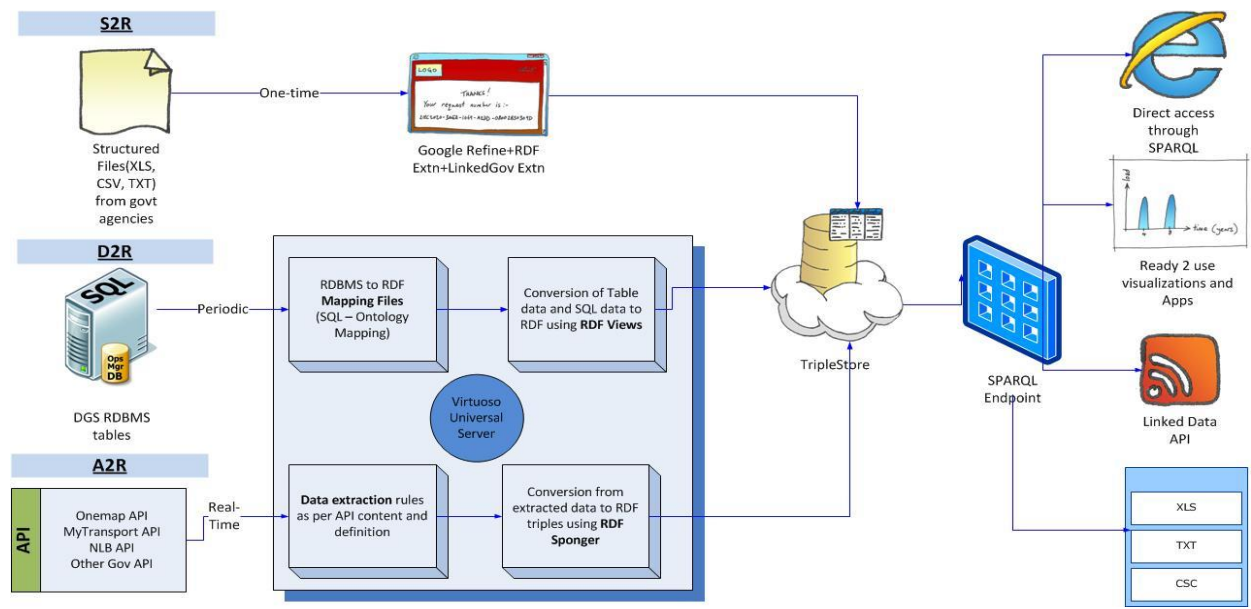

Figure 5. RDF Conversion Types and Endpoints for Singapore Open Data

### 4.5.1 Static Files Conversion (S2R)
Government agencies predominantly publish their data (about 80% of total government data sets) in static file format (Excel spreadsheets) in the form of tables or crosstabs. These files can be loaded into Google Refine (data transformation tool) that helps in curing, massaging and integration of fields in the data set. The project team used the URA Site Sales excel file for evaluating Google Refine.

### 4.5.2 Relational Database Conversion (D2R)
The Project team recommends the usage of STDTrip methodology devised by the Brazil Open Data team (Salas, Viterbo, Breitman, Casanova, 2011) before the D2R conversion process. STDTrip gives six distinct steps that aid in mapping RDBMS model to Vocabularies. The crux of the idea is to use ER diagram of DGS data model to get the corresponding classes and their properties. This process has some of its footing in Ontology Modeling step. The project team used the features of OpenLink's Virtuoso Server (Open source edition) extensively. One of the features is RDF Views that serves in D2R conversion. The mapping files generated from the STDTrip process is to be converted in RDF Views syntax for conversion. The conversion can be done on a daily basis (after the refresh of base DGS tables). The project team created a table directly in Universal Server for evaluating RDF Views and its mapping features.

### 4.5.3 Application Programming Interface Conversion (A2R)
RDF Sponger (another feature of Universal Server) was evaluated for A2R conversion. The A2R conversion is real-time. API services provided by SLA (OneMap API), LTA (myTransport API) and NLB (web services) can be authorized in the Universal Server with authentication keys followed by creation of external cartridges that are mapping files used to extract and map the contents of API output to corresponding classes and properties in the vocabularies used inside the data sets. Most of the APIs provide data in the form of JSON (OneMap) or XML (MyTransport) or plain text format. Separate cartridges are to be created for each of these content types. A sample OneMap API was used to test this functionally.

| Sub-step | Input | Process | Output |
|---|---|---|---|
| **STDTrip Process** | • Relational Model<br>• ER Model<br>• Identified Vocabularies | Seven step STDTrip process for generating mapping files | Mapping and Schema files |
| **S2R** | • Spreadsheets(xls, csv)<br>• Other text formats | • Cleansing<br>• Transformation | RDF Triples |
| **D2R** | • SQL of tables Mapping files | Conversion from table format to RDF | RDF Triples |
| **A2R** | • API URLs<br>• Mapping files | • Identification of API content type<br>• Conversion to RDF | RDF Triples |

**Table 13. Business Process Overview of RDF Creation Step**

### 4.5.4 Identified Issues

Issues in this step are mainly related to data quality, versioning, source system unavailability, URI management and non-adherence to mapping rules. Existing Linked Data literature on handling updates to existing data are not extensive and IDA would be best advised to devise a strategy that tests the prototype with multiple real-time updates (e.g. traffic related data of LTA). The real-time nature of A2R makes it completely dependent on source APIs. Absence of intimation about API outages can cause the system to return null or invalid results (if caching is enabled). Google Refine doesn't dereference the URI that it creates for each row in the source file. Dereferencing is to be done externally. Pubby provides this utility. Changes to SQL queries, updates to DB tables and changes to API output are to be indicated to the appropriate personnel before implementation. Mapping files are supposed to be updated before the aforementioned changes go into production.

### 4.5.5 Recommended Tools

Selection of the tool is largely dependent on the inter-operability features. The project team recommends Virtuoso Universal Server for its completeness for both D2R & A2R and Google Refine for S2R process because of its support for complex custom scripting that is needed while transforming data from its source format to the required format (for e.g. multidimensional nature of statistics related data provided by Department of Statistics).

## 4.6 STEP 6 - EXTERNAL LINKING

Linked Data is known for its prescription of four key principles. Even though, the adoption of fourth principle is optional for governments and enterprises, the web of data vision can be brought forth only by linking data sets across the globe (web of data depicted by LOD cloud). Current DGS system doesn't make use of external data or provide private firms data. Linked Data applications would be greatly benefited on the existence of links between government data sets and popular data. Project team has identified some scenarios of linking Singapore data sets with external data sets from an application development perspective as in Table 14.

---

1.) Wikipedia for use of textual content pertaining to places, events and entities in Singapore
2.) Geonames for use of spatial data
3.) Flickr wrapper for use of openly licensed images pertaining to places and events
4.) World Bank and CIA Factbook data sets to facilitate economic analysis of Singapore and other nations on common dimensions
5.) UN data sets such as FAO data sets for comparing import and export of commodities in Singapore
6.) Supreme Court database (SCDB) with DBpedia and News agency like LosAngeles Times and NewYork Times for analyzing the decisions made.

---

**Table 14. Examples of few potentially exploited external links**

Project team recommends data set owners to look at the open government data catalogue and datahub for potential data sets. The identified data sets and the government data sets can be compared using tools such as SILK and LIMES that automate the linking process between data sets. The linking is not contextual instead it is made between semantically related resources (use of the OWL property owl:sameas). Most of the links will be made to the data sets of DBpedia and Geonames as these services are regarded as good store hubs of general factual and spatial data respectively.

```
<http://data.gov.sg/location/bugis><owl:sameAs><http://www.dbpedia.org/resource/Bugis>
<http://data.gov.sg/race/chinese><owl:sameAs><http://www.dbpedia.org/resource/Malay_race>
```

**Table 15. Relating resources with ontology relationship**

| Sub-step | Input | Process | Output |
|---|---|---|---|
| External Linking | - Demographic and Spatial data sets (provided by DOS and SLA)<br>- Identified external data sets (DBpedia, Geonames, CIA Factbook) | Linking suggestions provided based on Similarity algorithms | Outbound links to external data sets |

**Table16. Business Process Overview of External Linking Step**

### 4.6.1 Identified Issues
The outbound links made to data sets outside of IDA's purview can be risky based on the situation of external source's data credibility being brought into question (e.g. Wikipedia is built on user content). Dead links are a vivid possibility during the change of resource URIs or system downtime. Application developers should take all these factors into consideration while using links involving data sources. There is an alternate solution of importing the RDF dump from the external data sets and provide information about data provenance within the data sets as a token of credit.

### 4.6.2 Recommended Tools
SILK and LIMES are industry standard tools that can automate the process of finding links based on similarity measures.

## 4.7 STEP 7 - DATASETS PUBLICATION

### 4.7.1 Triple Store
The generated RDF triples are to be stored in a Linked Data specific data structure called as Triple Store. Storage of triples can be centralized at DGS server or decentralized at individual agencies. A centralized storage is recommended to avoid different versions of truth. The Figure 6 in this section depicts a common triple store for storing the triples from the three conversion processes.

### 4.7.2 Metadata Publication
Metadata publication comes after the publication of actual data. Literature recommends the usage of VOID vocabulary for this purpose since it covers all facets such as general metadata, access privileges metadata and external links metadata. Metadata publication is a key differentiator of Linked Data on comparison to traditional Information systems as the triples are created with the purpose of reducing the learning time for the developer therefore accelerating the application development process.

### 4.7.3 SPARQL and Linked Data API
SPARQL is to Linked Data in the same vein as SQL is to Database system. RDF data can be queried only through SPARQL. The biggest advantage of SPARQL is the CONSTRUCT query that can generate new triples based on the characteristic of inference. For example, a set of triples can be generated between regions (Central, North) and cumulative site tender prices based on the common existence of 'Location' resource in two data sets, thus enables to move up and view a bigger picture.

SPARQL may not find acceptance from developers having experience in using SQL and REST APIs for development. An API wrapper called as Linked Data API can be stationed on top of SPARQL endpoint as a means of abstraction. Linked Data REST API translates the API call to SPARQL query, retrieves the resultant data from SPARQL and then translates the data to an appropriate output format such as JSON, XML, HTML or TEXT. Data extraction from Data.gov.uk is based on this concept. The URL http://education.data.gov.uk/doc/school/phase/secondary.json retrieves the list of secondary schools in UK in JSON format. A sequence based illustration of Linked Data API in DGS context is given in Figure 6. The use of LDA brings in obvious advantages as it simplifies data access and provides convenient output formats. This also means that the rules for conversion from API call to SPARQL are to be created. Project team foresees considerable time and effort to be invested in this activity as even a small change in API parameters would need a different SPARQL query.

### 4.7.4 Linked Data Hosting

Linked Data hosting is another option that can be considered by IDA if it aims to use third party platforms for data storage. Hosting in cloud brings in cost benefits, abstracts the complexity of maintaining servers & installing software and provides capability for instantly creating simple REST based APIs for data insertion, update, deletion & retrieval activities. The project team evaluated the Kasabi platform (powered by Talis) and found the interface to be intuitive. LOGD TWC portal provides this service for US government data sets. IDA can consider this option based on from the hosting vendor. UK Linked Data also makes use of Talis platform.

| Sub-step | Input | Process | Output |
|---|---|---|---|
| **Actual data publishing** | • S2R RDF triples<br>• D2R RDF triples<br>• A2R RDF triples | Data insertion through command line or in batch | Data storage in triple store |
| **Metadata publishing** | • Vocabulary classes<br>• Key Statistics from data set | Modeling in VOID | VOID triples |
| **SPARQL** | • SPARQL queries | Data retrieval or graph construction | Resultant RDF triples |
| **Linked Data API** | • REST based API Calls | • Conversion of API call to corresponding SPARQL query<br>• Conversion of resultant RDF triples to output format(e.g. JSON) | JSON/XML/HTML/TXT data |

**Table 17. Business Process Overview of Dataset Publication step**

### 4.7.5 Identified Issues

The use of named graphs is critical for IDA as updates to existing RDF triples are made in the triple store. A DML query can have unexpected results if the count of triples that are to be updated is unknown. Therefore, the use of named graphs is suggested. API based updates and deletion can be a preferred mode avoiding the complex nature of SPARQL queries. SPARQL does not currently support sub-queries and many other database operations that application developer make use of. The only alternative is to build complex SPARQL queries that do the same operation. This could lead to a long-winded process of updating queries for even small change requests. A possible solution can be SPARQL query creation/update template (in excel) that could aid the developers in accelerating the query creation process. Linked Data API with its simplicity actually eliminates the biggest advantage that Linked Data provides - inferencing through SPARQL CONSTRUCT query. Therefore, IDA is best advised to always maintain a public SPARQL endpoint to facilitate innovative querying. A security lapse or an unknown exploit in Linked Data hosting environment can lead to serious implications if the Linked Data is not open for public use. This risk always remains with external hosting. Evaluation of Vendor's backup measures is important before agreeing on the contract.

### 4.7.6 Recommended Tools
Virtuoso Universal Server's Triplestore is recommended for data publication and SPARQL endpoint. Puelia, PHP based Linked Data API library is recommended for API. If IDA plans to use Java for Linked Data API, Elda can be considered. Talis platform is recommended for Linked Data hosting.

## 4.8 STEP 8 - EXPLOITATION AND DISCOVERY
IDA's implementation of Linked Data over its existing data platform would be considered successful based on the number of new applications created with innovative ideas of using multi-agency data sets, number of API calls and number of new ideas implemented as applications. A government service provider would want the general public and enterprises to exploit its service offering and the same applies for its Linked Data offering. As a benchmark, the surge of applications after the release of UK government's Linked Data initiative has already been covered under the literature review section. IDA has started the ideas4apps initiative to look for new ideas that could be implemented as webapps or mobile apps build on top of government data. Such initiatives will have more impact with the data sets released in Linked Data format.

| Sub-step | Input | Process | Output |
| --- | --- | --- | --- |
| **Exploitation & Discovery** | • Linked Data<br>• Linked Data Existing Applications | • Gamification<br>• Crowdsourcing<br>• Converting Ideas to Apps<br>• Registering in public catalogues | • New Applications(e.g. Eagle Eye)<br>• External reference from LOD cloud |

**Table 18. Business Process Overview of Exploitation & Discovery Step**

The final implementation of the project team's use-case would be similar to that of in Figure 6. The middle box containing the site details for Pasir Ris is from the parent data set of URA. The box in the right contains demographic details of Pasir Ris, retrieved from the DOS data sets and the box in the left contains area facilities available in Pasir Ris, provided by SLA through OneMap API. The display of these two boxes with data coming from different agencies is made possible by using a common URI for Pasir Ris and by re-using this URI across three different data sources.

### 4.8.1 Identified Issues
There are no technical issues related to this step as it majorly related to publicity of the Linked Data system supported by IDA. Technical flaws, data quality issues and application level misrepresentations are the possible issues that could cascade from the earlier steps. Faceted browsing of Linked Data can cause serious performance issues as data retrieval is quite huge for this purpose. Public availability of this service can be restricted even though it is one of the best ways of slicing and dicing data for getting to the required point of satisfaction in data discovery.

### 4.8.2 Recommended Tools
Apache Jena Framework (initially developed by HP Labs) provides Java based APIs for dealing with most of the elements in the Semantic Web stack (RDF, Ontology and SPARQL). The project team used the SPARQL API for querying publicly available SPARQL endpoints. The framework is recommended based on its features and global usage. Linked Data browsers Sig.ma, Tabulator and OpenLink Data Explorer can be used over Singapore data sets for faceted browsing.

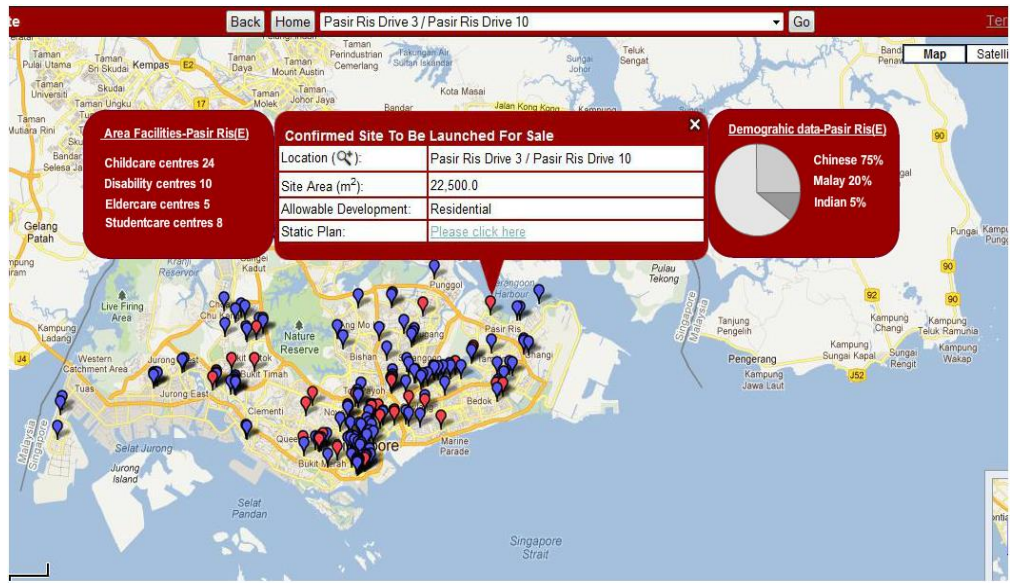

**Figure 6. Post Linked Data Migration – A snap shot of an application (After interlinking urban planning and population domains)**

## 5. CONCLUSION

This report's main contribution is its applicability to the Singapore Government data ecosystem thereby differentiating itself from other published Linked Data migration literature. iDA and government data agencies from other countries can use this report as a reference to plan their strategy for Linked Data implementation that seems imminent in the near future. The report shows the method of interlinking two data sets sourced by two different government agencies along with the steps involved in using the recommended tools (Google Refine, RDF Views and RDF Sponger) with these data sets. The identified issues and recommendations in each step will be highly useful during iDA's actual implementation. The work is incomplete in the view that a POC application was not built using the framework. Therefore, the project team foresees the possibility of unidentified data maintenance issues propping up during implementation. The experience of learning about new technologies, new methods of data representation, working through the case studies of other countries and contemplating on selection of best tools, was enriching for the project team. Most of the activities in the framework formulation involved systemic thinking.